# Electrooptical Graphene Plasmonic Logic Gates

Kelvin J. A. Ooi,[1,2,*] Hong Son Chu,[2] Ping Bai,[2] Lay Kee Ang[1]

[1]*Engineering Product Development, Singapore University of Technology and Design, 20 Dover Drive, Singapore 138682*

[2]*Electronics & Photonics Department, A*STAR Institute of High Performance Computing, 1 Fusionopolis Way, #16-16 Connexis, Singapore 138632*

[*]*Corresponding author: kelvin_ooi@sutd.edu.sg*

**Abstract:**

The versatile control of graphene's plasmonic modes via an external gate-voltage inspires us to design efficient electro-optical graphene plasmonic logic gates at the mid-infrared wavelengths. We show that these devices are superior to the conventional optical logic gates because the former possess both cutoff states and interferometric effects. Moreover, the designed 6 basic logic gates (i.e. NOR/AND, NAND/OR, XNOR/XOR) achieved not only ultra-compact size-lengths of less than $\lambda/28$ with respect to the operating wavelength of 10μm, but also a minimum extinction ratio of as high as 15dB. These graphene plasmonic logic gates are potential building blocks for future nanoscale mid-infrared photonic integrated circuits.

In recent years, the mid-infrared wavelength is attracting a lot of research interests due to potential device applications in areas ranging from free-space telecommunications, sensing and medicine, to infrared countermeasures and reconnaissance [1,2]. One of the components which might be found ubiquitously in those devices would be the electronic logic for signal processing. However, due to its architecture, the electronic logic inherently experiences undesirable effects such as latency and race-condition glitches, which might compromise the safety and reliability of those devices.

A better alternative would be to use the optical directed logic architecture [3], which has markedly-reduced gate propagation and state-change delays. The advantage of directed logic is prominently seen when logic gates are cascaded: because the gate and propagating signals are kept separate in directed logic, all gates can be switched simultaneously, very unlike electronic logic gates where their gate delays are cascaded. Consequently, the directed logic architecture can avoid many of the above-mentioned undesirable effects associated with the electronic logic. Recently, a range of silicon-on-insulator optical logic gates has been designed based on the directed logic architecture, which includes the Mach-Zehnder interferometer (MZI) and micro-ring resonator (MRR) types [4–10]. However, those logic gates have their own set of problems in certain logic-function implementations:

(i) In the AND/OR logic implementation for the MZI logic gate, there is a need for arbitrary definition of the ON/OFF threshold logic intensity levels [4,5]. This is due to the absence of cutoff states in the MZI logic gate architecture.

(ii) The MRR logic gate is superior in the AND/OR logic implementation due to having cutoff states [7,9,10]. However, difficulties arise in the XOR/XNOR logic implementation due to its lack of interferometric effects. As such, implementing these logics in MRR would either require additional switches and/or strategic placement of the switches [6–8], or using two wavelength signals [9].

In this Letter, we would like to introduce MZI-based graphene plasmonic logic gates as a common solution to the above-mentioned problems. Specifically, we want to show that the graphene plasmonic MZI logic gate does not face the problem in (i) because the plasmon propagation can be driven to the cutoff state, resulting in very distinct logic intensity levels for the ON/OFF states. Additionally, it is feasible to overcome the problem in (ii) due to its interferometric architecture. As an added benefit, graphene plasmonics operate efficiently in the mid-infrared spectrum, which coincides well with the operating spectrum of mid-infrared photonic devices.

Graphene is a two-dimensional, atomically-thick single layer of carbon atoms arranged in a honeycomb lattice, and exhibits many unique optical properties including the existence of plasmons [11]. Previously, we have discussed several interesting properties of graphene plasmons [12]. The doped graphene supports highly-confined plasmonic modes which can be broadly-tuned via chemical or electrostatic doping. This versatile control of graphene's plasmonic modes is attractive for highly-integrated mid-infrared active plasmonic devices in ultracompact platforms [13–15]. For example, in an active waveguide device, graphene plasmons can be excited and propagate along a doped graphene nanoribbon strip. The doping level and hence the propagation constants of the graphene plasmons can be easily modulated by a gate-voltage. In principle, the doped graphene can support the propagation of a wide spectrum of wavelengths, whereby the higher doping levels support the propagation of shorter wavelengths. For simplicity and technical feasibilities explained in our previous paper [12], in this paper the operating wavelength is chosen at 10μm together with the graphene nanoribbon width of 30nm, while the graphene nanoribbons are doped to 0.2eV and embedded in a silica substrate (these parameters gives a plasmon wavelength of ~80nm). The graphene nanoribbon relaxation time, which would determine its damping factor, is assumed to be 0.5ps throughout this paper. For simplicity, the edge effects of graphene nanoribbons, which might affect the Fermi level and conductivity of graphene [16], are assumed to be already taken into consideration in our analysis.

It is recently reported that plasmons can be coupled efficiently between two closely-spaced graphene sheets [15]. Thus, by using this coupling scheme, the graphene nanoribbon MZI logic gates can be easily constructed. The coupling between two identical graphene nanoribbons is enabled by the competition between the symmetric (β+) and antisymmetric (β-) modes of the graphene plasmons, with the coupling coefficient given by [15]:

$$C_g = \left| \frac{\beta_- - \beta_+}{2} \right| \quad (1)$$

The wave vectors β+ and β- are numerically computed by the mode solver in CST Microwave Studio 2012 [17], and the values are then used to find $C_g$. These 3 parameters are plotted in Fig. 1 as a function of the spatial separation t. There is clear indication from Fig. 1 that separation distance should be small to obtain a high plasmon coupling from one nanoribbon to another, which is similar to the result obtained in [15]. Meanwhile coupling lengths can be found through the equation [18]:

$$L_c = (m + \tfrac{1}{2})(\pi / \sqrt{2} C_g) \quad m = 0,1,2,... \quad (2)$$

which gives us $L_c \approx 37$nm for t = 10nm. Our simulation results showed that $L_c$ works well within the range of 15–60nm.

The basic switching unit (BSU) of our graphene plasmonic logic gate is shown in Fig. 2. Graphene plasmons travelling from the left input nanoribbon arm will split and couple to the top and bottom arms, and then recombine and couple back to the right output arm. In between the input and output arms, there is a separation gap $L_g$ (nominally chosen as 50nm) to halt the direct transmission of graphene plasmons from the input to the output of logic gates. The top and bottom arms are then modulated by connecting to an electrical source. The graphene nanoribbons are simply doped to the Fermi energy of $E_F$=0.2eV to allow a reasonable propagation distance at the operating wavelength of 10μm. During modulation, the doping level is reduced to <0.05eV by the electrical bias and thus driving the top and bottom arms to the cutoff state. However, it should be noted that all those device parameters can be changed to accommodate a variety of different logic gate functions in the following discussion.

Fig. 3 shows the realization of all 6 basic logic functions (i.e. NOR/AND, NAND/OR, XNOR/XOR) based on the graphene plasmonic MZI logic gate. We will first discuss the NOR/AND logic, which can be realized by cascading two BSUs in series, as shown in Fig. 3(a). For these logics, only one of the switching arm in each unit will be used, therefore the bottom arm will be removed or permanently biased to the cutoff state. For the NOR logic, the A and B input signals are connected to a pull-down circuit, so that under an unbiased condition, the plasmons on the graphene nanoribbon are coupled through the arms. For the AND logic, the pull-up circuit should be used to obtain the complementary effect.

Meanwhile, the NAND/OR logics only requires one BSU as shown in Fig. 3(b). The input signals for the NAND logic are connected to pull-down circuits, and pull-up circuits for the OR logic.

All the NOR/AND, NAND/OR logics operate under the cutoff-state switching implementation. However, for the XNOR/XOR logics, the interferometric switching should be used. Here, shown in Fig. 3(c), one BSU will be used but with a few modifications. Firstly, the top and bottom arms will be doped to a slightly higher chemical potential at 0.3eV, while the input/output arms are maintained at 0.2eV. Since the input/output and top/bottom arms have different chemical potentials, the coupling coefficient in Eq. (1) is modified as [19]:

$$C_{g2}^2 = C_g^2 + \left( \frac{\beta_0 - \beta_{1,2}}{2} \right)^2 \quad (3)$$

where $\beta_0$ is the wave vector of the input/output arm, and $\beta_1$ and $\beta_2$ are the wave vectors of the top and bottom arms respectively. Secondly, the applied bias should only reduce the arm doping level to 0.2eV, which still allows the plasmons to propagate but at another plasmon wavelength. When two arms have differing doping levels, a phase difference would be acquired, and this results in destructive interference occurring at some recombination point. We thus utilize this phase difference to construct the off-state for the XNOR/XOR logic. Thirdly, to achieve this destructive interference, $L_g$ should be modified to obtain a π-phase shift at the recombination point. The total length of the top/bottom arms should thus be written as:

$$L_g + 2L_c \approx m\pi / |k_1 - k_2| \quad m = 1,2,3,... \quad (4)$$

where $k_1$ is the effective propagation constant of the top arm, and $k_2$ is the effective propagation constant for the bottom arm. In our case, we found that the first two $L_g$ are 40nm and 75nm when the gate is embedded in silica. For the XNOR gate, both input signals are connected to pull-down circuits, while for the XOR gate, one arm is connected to a pull-down circuit, and another to a pull-up circuit.

In Table 1 we report our simulated extinction levels using the finite integral time domain method from CST Microwave Studio 2012 [17]. At ON states (logic level=1), the extinction levels (the optical attenuation of the plasmons) for all logics are at maximum –4dB. At OFF states (logic level=0), the extinction levels have a minimum of –19dB. This shows that the graphene plasmonic MZI logic gates are very efficient, having a minimum average extinction ratio between OFF/ON states of 15dB for all types of logic. In comparison to conventional MZI logic gates, especially for the NOR/AND and NAND/OR logics, they usually only have ~5dB extinction ratios [4,5]. We further show in Fig. 4 the electric-field maps for selected gates and input logics. The high performance of the logic gates is seen from the high contrast of the electric-field magnitudes between ON/OFF states. Furthermore, this high extinction ratio could be achieved for very compact logic devices, having overall device lengths of only from 150nm to 350nm. These values correspond to a length less than $\lambda/28$ or equivalent to 5 plasmon wavelengths.

The NOR/AND and NAND/OR logic gates have a linear frequency response since they operate with cut-off modes. In the case for the XNOR/XOR logic gate, however, due to the interferometric configuration, there is a resonant frequency response centered at 30THz (10μm) and a FWHM bandwidth of ~1THz as shown in Fig. 5. Thus, these graphene plasmonic logic gates are still potentially suitable for broadband applications.

In conclusion, we have designed all 6 basic logic functions by using the graphene plasmonic MZI implementation. We have shown that the graphene plasmonic logic gate is advantageous in possessing characteristics from both the cutoff state of MRR logic gates and also the interferometric effect of MZI logic gates, giving rise to its high extinction ratio. The logic devices can be made extremely compact together with high extinction ratios, typically less than 5 plasmon wavelengths. These design concepts are feasible to be scalable to a broad wavelength spectrum ranging from mid-infrared to THz wavelengths. These high performance graphene MZI logic gates may be potential building blocks for future nanoscale mid-infrared photonic integrated circuits.

This work is supported by the National Research Foundation Singapore under its Competitive Research Programme (CRP Award No. NRF-CRP 8-2011-07). KJAO and LKA would like to acknowledge the support of SUTD-MIT IDC grant (IDG21200106 and.IDD21200103).

**Table 1: Truth Table and corresponding extinction (dB) for graphene plasmonic MZI logic gates**

| Input | | Truth Table and Extinction (dB) | | | | | |
|---|---|---|---|---|---|---|---|
| A | B | NOR/AND* | (dB) | NAND/OR* | (dB) | XNOR/XOR* | (dB) |
| 0 | 0 | 1 | –2.7 | 1 | –1.5 | 1 | –3.8 |
| 0 | 1 | 0 | –31 | 1 | –3.7 | 0 | –19 |
| 1 | 0 | 0 | –30 | 1 | –3.7 | 0 | –19 |
| 1 | 1 | 0 | –47 | 0 | –31 | 1 | –1.8 |

*For AND and OR logics, outputs at the first and last rows should be inverted. For XOR logic, all outputs should be inverted.

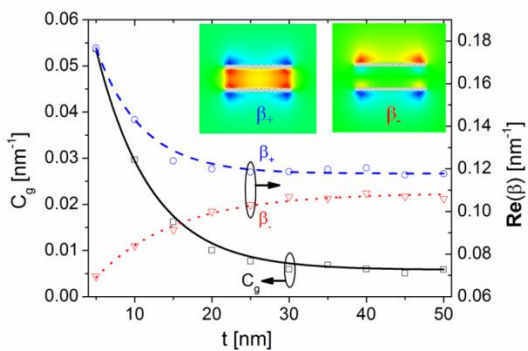

Fig. 1. Symmetric ($\beta+$) and antisymmetric ($\beta-$) graphene plasmon modes, and coupling coefficient $C_g$ between two graphene nanoribbons as a function of the spatial separation t. Inset: electric-field plots of the symmetric and antisymmetric graphene plasmon modes.

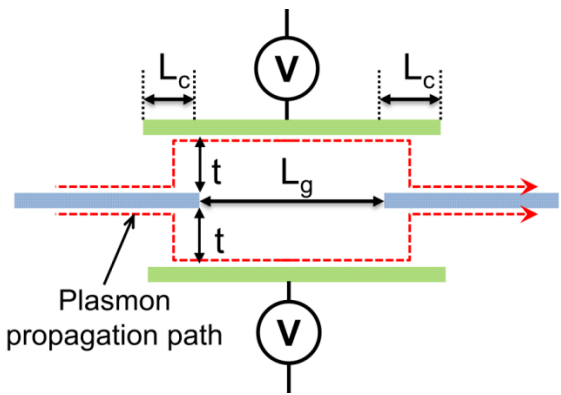

Fig. 2. Basic switching unit of the graphene plasmonic MZI logic gate.

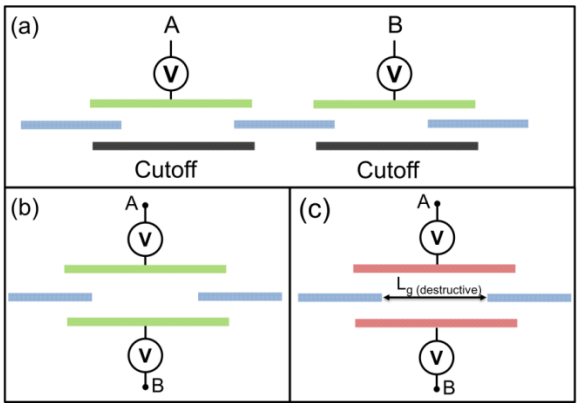

Fig. 3. Logic gate configuration for (a) NOR/AND, (b) NAND/OR and (c) XNOR/XOR logics.

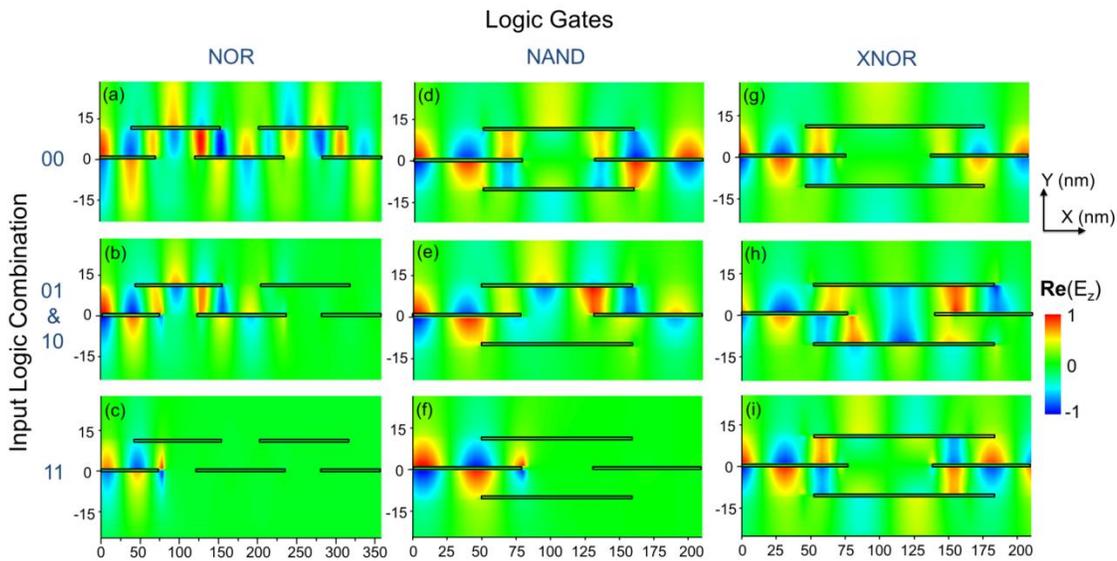

Fig. 4. Electric-field profiles for various graphene plasmonic logic gates and input logics. NOR gate under (a) 00, (b) 01&10 and (c) 11 input logic. NAND gate under (d) 00, (e) 01&10 and (f) 11 input logic. XNOR gate under (g) 00, (h) 01&10 and (i) 11 input logic.

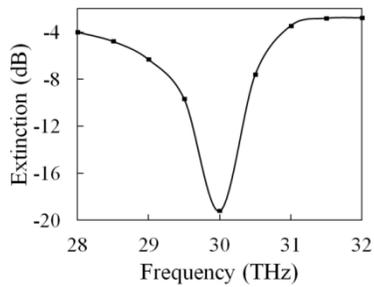

Fig. 5. Frequency response for the XNOR/XOR logic gate in the off-state.